\documentclass[twocolumn,preprintnumbers,amsmath,amssymb]{revtex4}
\usepackage{graphicx}
\usepackage{dcolumn}
\usepackage{bm}

\begin{document}
\title{Analytical device model for  graphene
bilayer field-effect transistors using weak nonlocality approximation
}

\author{ 
V.~Ryzhii\footnote{Electronic mail: v-ryzhii@u-aizu.ac.jp}$^{1,4}$,
M.~Ryzhii$^{1,4}$,
A.~Satou$^{2,4}$,
T. ~Otsuji$^{2,4}$,
and   V.~Mitin$^3$}
\affiliation{
$^1$Computational Nanoelectronics Laboratory, University of Aizu, 
Aizu-Wakamatsu  965-8580, Japan\\
$^2$Research Institute for Electrical Communication, Tohoku University, Sendai 980-8577, Japan\\
$^3$Department of Electrical Engineering,
University at Buffalo, State University of New York, NY 14260, USA\\ 
$^{4}$Japan Science and Technology Agency, CREST, Tokyo 107-0075, Japan\\
}

\begin{abstract}
We develop an analytical device model
for graphene bilayer
field-effect transistors (GBL-FETs) with the back and top gates. The model is based on the Boltzmann equation
for the electron transport and the Poisson equation in the weak nonlocality approximation
for the potential in the GBL-FET channel.
The potential distributions in the GBL-FET channel are found analytically.
The source-drain current in GBL-FETs and their transconductance
are expressed in terms of the geometrical parameters and applied voltages
by analytical formulas in the most important limiting cases.
These formulas explicitly account for the short-gate effect and the effect
of  drain-induced barrier lowering. The parameters characterizing the strength of these effects are derived. It is shown that the GBL-FET transconductance exhibits a pronounced
maximum as a function of the top-gate voltage swing. The interplay of
 the short-gate effect
and the electron collisions results in a nonmonotonic dependence of the transconductance
on the top-gate length. 
\end{abstract}

\maketitle

\section{Introduction}
Unique properties of graphene layers,  graphene nanoribbon arrays
and graphene bilayers
~\cite{1,2,3} as well as graphene nanomeshs~\cite{4}  make them promising
for different nanoelectronic device applications. 
The gapless energy spectrum of graphene layers allows to use them
in terahertz and midinfrared detectors and lasers~\cite{5,6,7,8,9}.
However, the gapless 
energy spectrum of GLs  is an obstacle for creating
transistor  digital circuits based on graphene field-effect transistors 
(G-FETs) 
due to relatively strong interband tunneling in the FET off-state~\cite{10,11}.
The reinstatement of the energy gap in graphene-based structures like
graphene nanoribbons, graphene nanomeshs, and graphene bilayers
appears to be unavoidable to fabricate FETs with a sufficiently large 
 on/off ratio.
Recently, the device dc and ac characteristics of 
graphene nanoribbon and graphene bilayer FETs 
(which are referred to as GNR-FETs and GBL-FETs, respectively)
were assessed both numerically and analytically~\cite{12,13,14,15,16,17,18,19}.
The device characteristics of  GNR-FETs
operating in  near ballistic and drift-diffusion regimes can be
calculated analogously with those of nanowire- and  carbon nanotube-FETs
(see, for instance ~\cite{20,21,22} and references therein).
The GBL-FET characteristics can, in principle, be found using 
the same approaches as those realized previously  for more customary FETs with
a two-dimensional electron system in the channel~\cite{23,24,25,26,27,28,29,30,31}.
However, some important features of GBL-FETs, in particular, the dependence
of both the electron density and the energy gap 
in different sections of the GBL-FET channel on the gate and drain voltages
should be considered~\cite{32,33,34}, as well as the ``short-gate'' effect and the drain-induced barrier lowering~\cite{29}.

In this paper, we use a  substantially generalized version of
the  GBL-FET analytical device 
model~\cite{17,18}
to calculate the characteristics of GBL-FETs 
(the 
threshold voltages, current-voltage characteristics, and transconductance)
in different regimes and
analyze the possibility of a significant
improvement of  the ultimate performance of these FETs
by shortening of the gate and decreasing of the gate layer thickness.
The device model under consideration, which presents the GBL-FET 
characteristics
in closed analytical form, allows a simple and clear evaluation
of the ultimate performance of GBL-FETs and their comparison. 
The following  effects 
are considered:
(a) Dependences of the electron density and energy gap in different sections
of the channel on the applied voltages and the inversion of the gated section charge;
(b) Degeneracy of the electron system, particularly, 
in the source and drain sections of the channel;
(c) The short-gate effect and the effect of drain-induced barrier lowering;
(d) Electron scattering in the channel.

Our model is based on the Boltzmann kinetic equation for the electron system
in the GBL-FET
 and the Poisson equation in the weak nonlocality approximation~\cite{35,36}. 
The use of the latter allows us to find the potential distributions
along the channel in the most interesting cases and obtain
the GBL-FET characteristics analytically.

The paper is organized as follows. In Sec.~II, the GBL-FET device model under consideration is presented and the features of  GBL-FET operation are discussed.
In Sec.~III, the main equations of the model are cited.  The general formulas for the source-drain current
and the GBL-FET transconductance  simplified for the limiting cases 
(far below the threshold, near threshold, and at low top-gate voltages corresponding
to the on-state)
are also presented. In this section, the  source-drain current
and the GBL-FET transconductance are expressed in terms of the Fermi energy in the source and drain contacts and the height of the potential barrier in the channel.
To find the barrier height,  the Poisson equation is solved for different 
limiting cases in Sec.~IV. The obtained potential distributions are used for 
the derivation of the explicit formulas for source-drain current and the transconductance
as functions of the  applied voltages and geometrical parameters.
Section~V deals with a brief discussion  of  some effects 
(role of the device geometry, electron scattering, charge inversion in the channel, and interband tunneling), which influence the GBL-FET
characteristics.
In Sec.~VI,  we draw the main conclusions. Some reference data related to
the voltage dependences of the Fermi energy and the energy gap in different sections of the GBL-FET channel are singled out to the Appendix.

\section{Device model and features of operation}
\begin{figure}[t]
\begin{center}
\includegraphics[width=7.0cm]{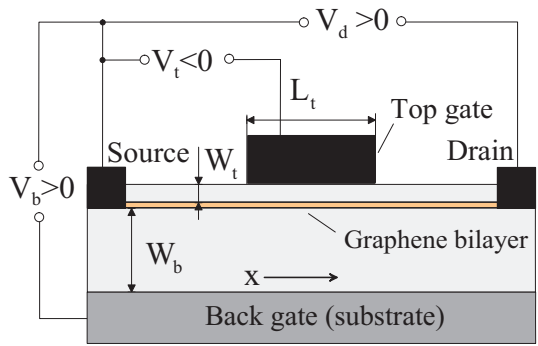}
\end{center}
\label{fig1}
\caption{Schematic view of the GBL-FET structure. 
}
\end{figure}

We consider  a GBL-FET with the structure shown in Fig.~1.
It is assumed that the  back gate, which is positively biased
by the pertinent voltage $V_b > 0$, provides the formation of the electron channel 
in  the GBL  between the Ohmic source and drain contacts. 
A relatively short top gate
serves to control the source-drain current by forming the potential barrier
(its height $\Delta_m$ depends on the top gate voltage $V_t$ and other voltages)
for the electrons propagating between the contacts.

We shall assume that the GBL-FETs
under the conditions when the electron systems in the source
and drain sections  are  degenerate, i.e., $\varepsilon_F
\gg k_BT$. This implies that the back gate voltage
is sufficiently high to induce necessary electron density 
in the source and drain sections.

In the GBL-FET the energy gap is electrically induced by 
the back gate voltage~\cite{32,33,34} (see also~\cite{18}). Thus in GBL-FETs, 
the back gate plays the dual role: it provides the formation of the electron channel and 
the energy gap. Since the electric field component
directed perpendicular to the GBL plane in the channel section below the top
gate (gated section) is determined by both $V_b$ and $V_t$, the energy gap can be different
in different sections of the GBL channel: $E_{g,s}$ (source section), 
$E_g$ (gated section), and $E_{g,d}$
(drain section)~\cite{17,18}.
At sufficiently strong top-gate voltage ($V_t < V_{th} < 0$, where $V_{th}$ is the threshold voltage), 
the gated section becomes depleted.
Since the energy gaps in  GBLs  are in reality not particularly wide,
at further moderate increase in $|V_t|$,  
the gated section of the channel
becomes filled with holes (inversion of the charge in the gated section)
if $V_t < V_{in} < V_{th}$, where $V_{in}$ is the inversion voltage. 
As a result, the  GBL-FETs with moderate energy gap are characterized
by the threshold and inversion voltages: $V_{th} < 0$ and $V_{in} < 0$.
The explicit formulas for  $V_{th}$ and $V_{in}$
shall be given in the following.
The cases $V_t = V_{th}$  and
 $V_t = V_{in}$ correspond to 
the alignment of the Fermi level in the source section of the channel with the conduction band bottom and the valence band top, respectively, 
in the gated section.


\section{Main equations of the model}

Due to relatively high energy of optical phonons in graphene,
the electron scattering in the GBL-FET channel is  primarily due to
disorder and acoustic phonons.
Considering such quasielastic scattering,
the quasiclassical Boltzmann kinetic equation governing 
the steady state electron distribution function
$f_{\bf p} = f_{\bf p}(x)$ in the gated section of the channel can
be presented as 
\begin{equation}\label{eq1}
v_x\frac{\partial f_{\bf p}}{\partial x} + e
\frac{\partial \varphi}{\partial x}\frac{\partial f_{\bf p}}{\partial p_x}
= \int\, d^2{\bf q}w(q)(f_{{\bf p}+ {\bf q}} - f_{\bf p})
\delta (\varepsilon_{{\bf p}+ {\bf q}} -  \varepsilon_{\bf p}).
\end{equation}
Here  $e = |e|$ is the electron charge, $\varepsilon_{\bf p} = p^2/2m$, 
$m\sim 0.04m_0$ is the electron effective mass in GBL
($m_0$ is the bare electron mass), ${\bf p} = (p_x,p_y)$ is the electron momentum in the GBL plane ($z = 0$), $w(q)$ is the probability of the electron 
scattering on disorder and acoustic phonons with the variation
of the electron momentum by quantity
${\bf q} = (q_x,q_y)$, $v_x = p_x$,  and axis $x$ is directed
in this plane (from the source contact to the drain contact, i.e., in the direction of the current). For simplicity, we disregard the effect of 
``Mexican hat''(see, for instance, Ref.~\cite{34}) and a deviation of
the real energy spectrum in the GBL from the parabolic one (the latter can
be marked in the source and drain sections with relatively high Fermi energies).
One of the potential advantages of GBL-FETs is the possibility
of ballistic transport even if the top-gate length $L_t$ is not small.
In such GBL-FETs, one can neglect the right-hand side term in Eq.~(1).

As in ~Ref.~\cite{10,15,34,36},
we use the following equation
for  the electric potential  $\varphi = \varphi(x) = \psi (x,z)|_{z = 0}$
in the GBL plane:
\begin{equation}\label{eq2}
\frac{(W_b + W_t)}{3}\frac{\partial^2 \varphi}{\partial x^2}
- \frac{\varphi - V_b}{W_b}- \frac{\varphi - V_t}{W_t}
= \frac{4\pi\,e}{k} (\Sigma_{-} - \Sigma_{+}).
\end{equation}
Here, $\Sigma_{-}$ and $\Sigma_+$ are the electron and hole sheet densities
in the channel, respectively,  $k$ is the dielectric constant of the 
layers between the GBL and the gates, and $W_b$ and $W_t$ are the thicknesses
of this layers. In the following, we put $W_b = W_t =W$ (except Sec.~V).
Equation~(2) is a consequence of the two-dimensional Poisson equation
for the electric potential $\psi (x,z)$ in the GBL-FET gated section
($-L_t/2 \leq x \leq L_t/2$ and $-W_b \leq z \leq W_t$, where $L_t$ is the length of the top gate)
in the weak nonlocality approximation~\cite{35}. 
This equation provides the potential distributions, 
which can be obtained from
the two-dimensional Poisson equation  by expansion in powers of the parameter $\delta = (W_b^3 + W_t^3)/15(W_b + W_t)/{\cal L}^2 
= W^2/15{\cal L}^2$, where ${\cal L}$
is the characteristic scale of the lateral inhomogeneities 
(in the $x$-direction) assuming that $\delta \ll 1$, i.e.,
${\cal L}$ is not too small.
The lowest approximation in such 
an expansion leads to
the Shockley's gradual channel approximation, in which the first term in
the left side of Eq.~(2) is neglected~\cite{37,38}. 
The factor $1/3$ appeared due to features of 
the Green function of the Laplace operator in the case of the geometry 
under consideration.

The boundary conditions for Eqs.~(1) and (2)
are presented
as
\begin{equation}\label{eq3}
f_{\bf p}\biggr|_{p_x \geq 0, x = -L_t/2} = f_{s,{\bf p}}, \qquad
f_{\bf p}\biggr|_{p_x \leq 0, x = L_t/2} = f_{d,{\bf p}},
\end{equation}
\begin{equation}\label{eq4}
\varphi\biggl|_{x = -L_t/2} = 0, 
\qquad \varphi\biggl|_{x = L_t/2} = V_d + (\varepsilon_{F,d} - \varepsilon_{F,s})/e
= V_d^*,
\end{equation}
where 
$f_{s,{\bf p}}$ and $f_{d,{\bf p}}$ are the electron distribution
functions in the source and drain sections of the
channel.
The functions $f_{s,{\bf p}}$ and $f_{d,{\bf p}}$  are the Fermi distribution functions with
the Fermi energies $\varepsilon_{F,s}$ and  $\varepsilon_{F,d}$, which are determined by the back gate and drain voltages, $V_b$ and $V_d$~\cite{17,18} (see also the Appendix):
\begin{equation}\label{eq5}
\varepsilon_{F,s} \simeq eV_b\frac{b}{(1 + b)},
\qquad  \varepsilon_{F,d} \simeq e(V_b - V_d)\frac{b}{(1 + b)},
\end{equation}
where $b = a_B/8W$, $a_B = k\hbar^2/me^2$ is the Bohr radius, and
 $\hbar$ is the reduced Planck constant.
In the following, we shall  assume that $b  \ll 1$, so that
$\varepsilon_{F,s} \simeq beV_b$ and $\varepsilon_{F,d} \simeq be(V_b - V_d)$.
In particular, if $a_B = 4$~nm (GBL on SiO$_2$) 
 and $W = 10$~nm, one obtains
$b \simeq 0.05$.
Due to a smallness of $b$, we shall 
disregard a distinction between $V_d^*$ and $V_d$
because $V_d - V_d^* \simeq bV_d \ll V_d$ (as shown in the Appendix). 
Restricting ourselves by the consideration
of GBL-FETs operation at  not too high  drain voltages, 
we also  neglect the difference in
the Fermi energies in the source and drain sections, 
i.e., put $\varepsilon_{F,d} \simeq \varepsilon_{F,s} =
\varepsilon_F$. 

The source-drain dc current density (current per 
unit length in the direction perpendicular its flow)
 can be calculated using the following formulae:
$$
J = \frac{4e}{(2\pi\hbar)^2}\int\,d^2{\bf p}v_xf_{\bf p}
$$
\begin{equation}\label{eq6}
= \frac{e}{\pi^2\hbar^2}\int_{-\infty}^{\infty}dp_y\int_0^{\infty}dp_x\,
v_x
(f_{\bf p} - f_{\bf -p}).
\end{equation}
In this case,
Eq.~(1) with  boundary conditions~(3) yield

\begin{equation}\label{eq7}
f_{\bf p} - f_{\bf -p}\simeq \frac{\Theta(p_x^2/2m + e\varphi) - \Theta(p_x^2/2m + e\varphi - eV_d)}{1 + \exp[(p^2/2m + e\varphi - \varepsilon_F)/k_BT]},
\end{equation}
where  $T$ is the temperature,  $k_B$
is the Boltzmann constant, and 
$\Theta(\varepsilon)$ is the unity step function.
Using Eqs.~(6) and  (7), 
we obtain
\begin{widetext}
$$
J 
= \frac{e}{\pi^2\hbar^2}\int_{-\infty}^{\infty}dp_y
\int_{\Delta_m}^{\infty}
d\xi\,\biggl\{\frac{1}{1 + \exp[(p_y^2/2m + \xi - \varepsilon_F)/k_BT]}
- \frac{1}{1 + \exp[(p_y^2/2m 
+ \xi - \varepsilon_F + eV_d)/k_BT]}\biggr\}
$$
\begin{equation}\label{eq8}-
= \frac{ek_BT}{\pi^2\hbar^2}\int_{-\infty}^{\infty}dp_y
\biggl\{
\ln\biggl[\exp\biggl(\frac{\varepsilon_F -p_y^2/2m - \Delta_m}{k_BT}\biggr) +  1\biggr] - 
\ln\biggl[\exp\biggl(\frac{\varepsilon_F -p_y^2/2m - \Delta_m - eV_d}{k_BT}\biggr) +  1\biggr]
\biggr\}.
\end{equation}
\end{widetext}
Equation~(8) can be presented in the following form:
$$
J = J_0\,\int_{0}^{\infty}dz
\biggl\{
\ln\biggl[\exp(\delta_m - z^2) +  1\biggr]
$$
\begin{equation}\label{eq9}
-
\ln\biggl[\exp(\delta_m - U_d - z^2) +  1\biggr] 
\biggr\}.
\end{equation}
Here (see, for instance, Ref.~\cite{23})
\begin{equation}\label{eq10}
J_0 = \frac{2\sqrt{2m}e(k_BT)^{3/2}}{\pi^2\hbar^2},
\end{equation} 
is the characteristic current density, and 
$\delta_m = (\varepsilon_F - \Delta_m)/k_BT$, and $U_d = eV_d/k_BT$
are the normalized voltage swing and drain voltage, respectively.
At $m = 4\times 10^{-29}$~g and $T = 300$~K, $J_0 \simeq 2.443$~A/cm.

Figure~2 shows the dependences of the  source-drain current $J$ normalized 
by the value $J_0$
as a function of the $U_d$ calculated using Eq.~(9) for different
values of $\delta_m$.

The GBL-FET transconductance $g$ is defined as
\begin{equation}\label{eq11}
g = \frac{\partial J}{\partial V_t}.
\end{equation}
Equations (9) and (11) yield
$$
g = J_0 
\,\int_{0}^{\infty}dz
\biggl\{\biggl[\exp(z^2 - \delta_m) +  1\biggr]^{-1}
$$
\begin{equation}\label{eq12}
-
\biggl[\exp(z^2 - \delta_m + U_d) +  1\biggr]^{-1} 
\biggr\}\,\biggl(-\frac{\partial\,\delta_m}{\partial\,V_t}\biggr).
\end{equation}
\begin{figure}[t]
\begin{center}
\includegraphics[width=7.0cm]{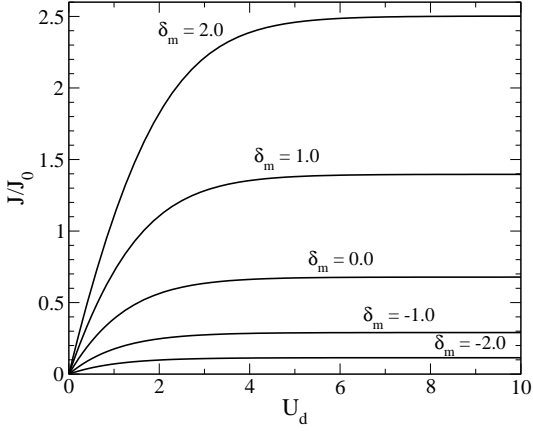}
\end{center}
\label{fig2}
\caption{Normalized source-drain current $J/J_0$ versus normalized drain voltage $U_d$
at different values of normalized gate-voltage swing $\delta_m$. 
}
\end{figure}

The obtained formulas for the source-drain current and transconductance
can be simplified in the following limiting cases: 

{\it A. High top-gate voltages}.
At high top-gate voltages, which correspond to the sub-threshold voltage range,
 the barrier height exceeds the Fermi energy
($\Delta_m \gg \varepsilon_F$),
so that  $ \delta_m \gg 1$. In this case (the electron system in the gated section 
is nondegenerate), using  Eqs.~(9) and (12), we obtain

\begin{equation}\label{eq13}
J = \frac{\sqrt{\pi}}{2}J_0\,\exp\biggl(
 \frac{\varepsilon_F - \Delta_m}{k_BT}\biggr)
\biggl[1 - 
\exp\biggl(- \frac{eV_d}{k_BT}\biggr)\biggr],
\end{equation}
\begin{equation}\label{eq14}
g = \frac{\sqrt{\pi}}{2}\frac{J_0}{k_BT} \,\exp\biggl(
 \frac{\varepsilon_F - \Delta_m}{k_BT}\biggr)
\biggl[1 - 
\exp\biggl(- \frac{eV_d}{k_BT}\biggr)\biggr]
\biggl(-\frac{\partial\,\Delta_m}{\partial\,V_t}\biggr).
\end{equation}

{\it B. Near threshold top-gate voltages.}
In this case, $\Delta_m \gtrsim \varepsilon_F$, i.e., $|\delta_m| \lesssim 1$,
Eqs.~(9) and (12) yield

\begin{equation}\label{eq15}
J \simeq J_0\,\frac{eV_d}{k_BT}
\biggl[\zeta_1  + \zeta_2\biggl(\frac{\varepsilon_F - \Delta_m}{k_BT}\biggr)
\biggr]\biggl[1 - 
\exp\biggl(- \frac{eV_d}{k_BT}\biggr)\biggr],
\end{equation}
\begin{equation}\label{eq16}
g \simeq \frac{J_0eV_d}{(k_BT)^2} \zeta_2
\biggl(-\frac{\partial\,\Delta_m}{\partial\,V_t}\biggr)
\end{equation}
at low drain voltages $eV_d \lesssim k_BT$ ($U_d \lesssim 1$), and

\begin{equation}\label{eq17}
J \simeq J_0\,
\biggl[\zeta_0  + \zeta_1\biggl(\frac{\varepsilon_F - \Delta_m}{k_BT}\biggr)
\biggr],
\end{equation}
\begin{equation}\label{eq18}
g \simeq \frac{J_0}{k_BT} \zeta_1
\biggl(-\frac{\partial\,\Delta_m}{\partial\,V_t}\biggr)
\end{equation}
at high drain voltages $eV_d \gg k_BT$ ($U_d \gg 1$).
Here, 
$\zeta_0 = \int_0^{\infty}d\xi\ln[\exp(-\xi^2) + 1] \simeq 0.678$, 
$\zeta_1 = \int_0^{\infty}d\xi/[\exp(\xi^2) + 1] \simeq 0.536$, 
and
$\zeta_2 =  \int_0^{\infty}d\xi\exp(\xi^2)/[\exp(\xi^2) + 1]^2$.
In the limit $\varepsilon_F = \Delta_m $,  Eqs.~(13) and (14)
 provide the values $J$ and $g$
close to those obtained from Eqs.~(17) and (18),
 which are rigorous in such a limit.

{\it C. Low top-gate voltages.}
At low top-gate voltages,  $\Delta_m < \varepsilon_F$,
from Eqs.~(9)

$$
J \simeq \frac{2}{3}I_0[(\varepsilon_F - \Delta_m)^{3/2}\Theta(\varepsilon_F - \Delta_m)
$$
\begin{equation}\label{eq19}
 - 
(\varepsilon_F- \Delta_m - eV_d)^{3/2} \Theta(\varepsilon_F- \Delta_m - eV_d)].
\end{equation}
%
$$
g \simeq I_0[(\varepsilon_F - \Delta_m)^{1/2}\Theta(\varepsilon_F - \Delta_m)
$$
\begin{equation}\label{eq20}
 - 
(\varepsilon_F- \Delta_m - eV_d)^{1/2} 
\Theta(\varepsilon_F - \Delta_m - eV_d)]
\biggl(-\frac{\partial\,\Delta_m}{\partial\,V_t}\biggr).
\end{equation}
Here, 
\begin{equation}\label{eq21}
I_0 = \frac{2\sqrt{2m}e}{\pi^2\hbar^2},
\end{equation}
with
$J_0 = I_0(k_BT)^{3/2}$.

Using Eqs.~(19) and (20), one obtains 
\begin{equation}\label{eq22}
J \simeq I_0\,eV_d\sqrt{(\varepsilon_F - \Delta_m)} \simeq  
I_0\,eV_d \sqrt{\varepsilon_F}
\biggl(1  - \frac{\Delta_m}{2\varepsilon_F}\biggr),
\end{equation}
\begin{equation}\label{eq23}
g \simeq \frac{1}{2}I_0\,
\frac{eV_d}{\sqrt{\varepsilon_F}}
\biggl(-\frac{\partial\,\Delta_m}{\partial\,V_t}\biggr)
\end{equation}
at  $eV_d \ll  \varepsilon_F - \Delta_m$, 
and
\begin{equation}\label{eq24}
J =  \frac{2}{3}I_0(\varepsilon_F - \Delta_m)^{3/2}
\simeq \frac{2}{3}I_0\varepsilon_F^{3/2}
\biggl(1 - \frac{3\Delta_m}{2\varepsilon_F}\biggr),
\end{equation}
\begin{equation}\label{eq25}
g \simeq I_0\,
\sqrt{(\varepsilon_F - \Delta_m)}
\biggl(-\frac{\partial\,\Delta_m}{\partial\,V_t}\biggr)
\end{equation}
at $eV_d \gg  \varepsilon_F - \Delta_m$.

The dependences shown in Fig.~2 describe implicitly the dependences of $J$
calculated using the universal Eq.~(9) on
the back-gate, top-gate, and drain  voltages as well as on the geometrical
parameters.
Equations~ (13) -  (25) provide these  dependences 
in most interesting  limits. 
However,
to obtain the explicit  formulas for $J$ as well as for $g$, one needs
to determine the dependences of the barrier height $\Delta_m$
on all voltages and geometrical parameters. 
Since the electron densities in the gated section
in the limiting cases under consideration are different, the screening 
abilities of the electron system in this section and the potential
distributions are also different. The latter
leads to different $\Delta_m$ vs $V_t$ relations.

\section{Potential distributions, source-drain current, and transconductance}

To obtain the explicit  dependences of the source-drain current and the transconductance
on the gate voltages $V_b$ and $V_t$ as well as on the drain voltage $V_d$, one needs
to find the relationship between the barrier height $\Delta_m$ and these voltages.
This necessitates the calculations of the potential distribution in the channel.
The latter can be found from Eq.~(2)
in an analytical form in the following limiting cases.

\subsection{ High top-gate voltages - sub-threshold voltage range
 ($\Delta_m \gg \varepsilon_F$).}

When the barrier height $\Delta_m$ exceeds the Fermi energy $\varepsilon_F$, 
the  electron density is low in the gated section and,
hence, one can disregard
the contribution of the electron charge in this section.
In such a  limit, we arrive at the following equation for the potential:
\begin{equation}\label{eq26}
\frac{d^2 \varphi}{d x^2} -
\frac{\varphi}{\Lambda_0^2}
=  \frac{F_0}{\Lambda_0^2},
\end{equation}
where  $\Lambda_0 = \sqrt{2/3}\,W$ 
and   $F_0 =
-(V_b + V_t)/2$.
Solving Eq.~(26) considering boundary conditions~(4),
for the case of high top-gate voltages we obtain 
\begin{equation}\label{eq27}
\varphi =  F_0
\biggl[\frac{\cosh(x/\Lambda_0)}
{\cosh(L_t/2\Lambda_0)} - 1 \biggr] +  V_d\,\frac{\sinh[(2x + L_t)/2\Lambda_0]}
{\sinh(L_t/\Lambda_0)}.
\end{equation}
Limiting our consideration by the GBL-FETs with not too short top gate
($L_t \gg W$), Eq.~(27) can  be presented as
\begin{equation}\label{eq28}
\varphi\simeq - F_0
\biggl[1 - 2\exp\biggl(-\frac{L_t}{2\Lambda_0}\biggr)\cosh\biggl(\frac{x}
{\Lambda}\biggr)
\biggr] + V_d\exp\biggl(-\frac{L_t}{2\Lambda_0}\biggr)
\exp\biggl(\frac{x}
{\Lambda_0}\biggr).
\end{equation}
Equation~(28) yields
\begin{equation}\label{eq29}
\Delta_m \simeq eF_0
\biggl(1 - \frac{1}{\eta_0}\biggr)
-  \frac{eV_d}{2\eta_0} =  - \frac{e(V_b + V_t)}{2}
\biggl(1 - \frac{1}{\eta_0}\biggr)
-  \frac{eV_d}{2\eta_0},
\end{equation}
where $\eta_0 = \exp (L_t/2\Lambda_0)/2$.
Simultaneously for  the position of the barrier top one obtains
\begin{equation}\label{eq30}
x_m = -\frac{\Lambda_0}{2}\ln\biggl(1 + \frac{V_d}{F_0}\biggr) =  
-\frac{\Lambda_0}{2}\ln\biggl(1 - \frac{2V_d}{V_b + V_t}\biggr).
\end{equation}

The terms in the right-hand side of Eq.~(29)
containing parameter $\eta_0$
 reflect 
the effect of the top-gate geometry (finiteness of its length).
This  effect is weakened with increasing
top barrier length $L_t$. 
The effect of  drain-induced barrier lowering 
in the case under consideration is described by the last term
in the right-hand side of Eq.~(29).

Equation~(29) yields $(\partial \Delta_m/\partial V_t) = - 
(e/2)(1- \eta_0^{-1})$.
Invoking Eqs.~(13) and  (14), we obtain

$$
J = \frac{\sqrt{\pi}}{2}J_0\,\exp\biggl[
 \frac{e(V_t - V_{th})}{2k_BT}\biggl(1 - \frac{1}{\eta_0}\biggr)\biggr]
$$
\begin{equation}\label{eq31}
\times\biggl[1 - 
\exp\biggl(- \frac{eV_d}{k_BT}\biggr)\biggr]\exp\biggl(\frac{eV_d}{2\eta_0k_BT}\biggr),
\end{equation}

$$
g \simeq \frac{\sqrt{\pi}}{4}\frac{eJ_0}{k_BT} \,\exp\biggl[
 \frac{e(V_t - V_{th})}{2k_BT}\biggl(1 - \frac{1}{\eta_0}\biggr)
\biggr]
$$
\begin{equation}\label{eq32}
\times\biggl[1 - 
\exp\biggl(- \frac{eV_d}{k_BT}\biggr)\biggr]\exp\biggl(\frac{eV_d}{2\eta_0k_BT}\biggr).
\end{equation}
Here, 
$V_{th} = -[1 + 2b/(1 - \eta_0^{-1})]V_b \simeq -(1 + 2b)V_b$. 
The rightmost factors in the right-hand sides of
Eqs.~(31) and (32), associated with the  effect  of drain-induced barrier lowering, 
lead to  an increase
in $g$  with increasing $V_d$ not only at $eV_d \sim k_BT$ but
at $eV_d \gg k_BT$: $g \propto \exp (eV_d/2\eta_0k_BT)$.
One can see that in the range of the top-gate voltages under consideration,
the GBL-FET 
transconductance exponentially decreases with increasing $|V_t + V_b|$
and 
\begin{equation}\label{eq33}
g \simeq  \frac{Je}{2k_BT} \ll g_0 =  \frac{\sqrt{\pi}}{4}\frac{eJ_0}{k_BT},
\end{equation}
where at $T = 300$~K 
the characteristic value of the transconductance $g_0 \simeq 4330$~mS/mm.

\subsection{Near threshold top-gate voltages ($\Delta_m \gtrsim \varepsilon_F$).}

At $ eV_d \lesssim k_BT \ll \varepsilon_F$, 
taking into account that the electron distribution  is characterized by
the equilibrium Fermi distribution function,
the electron density in the gated section can be presented in the following form:
\begin{equation}\label{eq34}
\Sigma \simeq \frac{2m}{\pi\hbar^2}(\varepsilon_F + e\varphi).
\end{equation}
Considering this, we reduce Eq.~(2) to
\begin{equation}\label{eq35}
\frac{d^2 \varphi}{d x^2} -
\frac{\varphi}{\Lambda^2}
=  \frac{F}{\Lambda^2}.
\end{equation}
Here, 
$$
\Lambda = \sqrt{\frac{a_BW}{12(1  + 2b)}} \simeq \sqrt{\frac{a_BW}{12}}
=  W\sqrt{\frac{2}{3}b},
$$
$$
F = \frac{[\varepsilon_F/e - b(V_b + V_t)]}{(1 + 2b)} 
\simeq  - b(bV_b + V_t) \simeq -bV_t,
$$
so that $\Lambda/\Lambda_0 \simeq \sqrt{b} < 1$.
The solution of Eq.~(35) with boundary condition~(4) is given by
$$
\varphi =  F
\biggl[\frac{\cosh(x/\Lambda)}
{\cosh(L_t/2\Lambda)} - 1 \biggr] + V_d\,\frac{\sinh[(2x + L_t)/2\Lambda]}
{\sinh(L_t/\Lambda)}
$$
$$
\simeq  - F
\biggl[1 - 2\exp\biggl(-\frac{L_t}{2\Lambda}\biggr)\cosh\biggl(\frac{x}
{\Lambda}\biggr)
\biggr] 
$$
\begin{equation}\label{eq37}
+ V_d\exp\biggl(-\frac{L_t}{2\Lambda}\biggr)
\exp\biggl(\frac{x}
{\Lambda}\biggr).
\end{equation}
From Eq.~(36) we obtain
$$
\Delta_m \simeq eF
\biggl(1 - \frac{1}{\eta}\biggr)
-  \frac{eV_d}{2\eta} 
$$
\begin{equation}\label{eq37}
\simeq   [\varepsilon_F - eb(V_b + V_t)]
\biggl(1 - \frac{1}{\eta}\biggr)
-  \frac{eV_d}{2\eta} \simeq   - ebV_t
\biggl(1 - \frac{1}{\eta}\biggr)
-  \frac{eV_d}{2\eta},
\end{equation}
where $\eta = \exp (L_t/2\Lambda)/2$, and the position of the barrier top
is 
\begin{equation}\label{eq38}
x_m \simeq 
-\frac{\Lambda}{2}\frac{V_d}{F} \simeq -\frac{\Lambda}{2b}\frac{V_d}{V_b}.
\end{equation}
Since $\Lambda < \Lambda_0$, one obtains $\eta \gg \eta_0$, and
the terms in Eq.~(37) containing parameter $\eta$
can be disregarded. This implies that the effects of  top-gate geometry and
 drain-induced barrier lowering are much weaker (negligible)
in the case of  the top-gate voltages
in question in comparison with the case of  high top-gate voltages.

Substituting $\Delta_m$ from Eq.~(37) into Eq.~(16), for
low drain voltages we arrive at

\begin{equation}\label{eq39}
g \simeq \frac{J_0e^2V_d\zeta_2}{(k_BT)^2}\biggl(1 -\frac{1}{\eta}\biggr)b. 
\end{equation}
%

\begin{figure*}[t]
\begin{center}
\includegraphics[width=12.0cm]{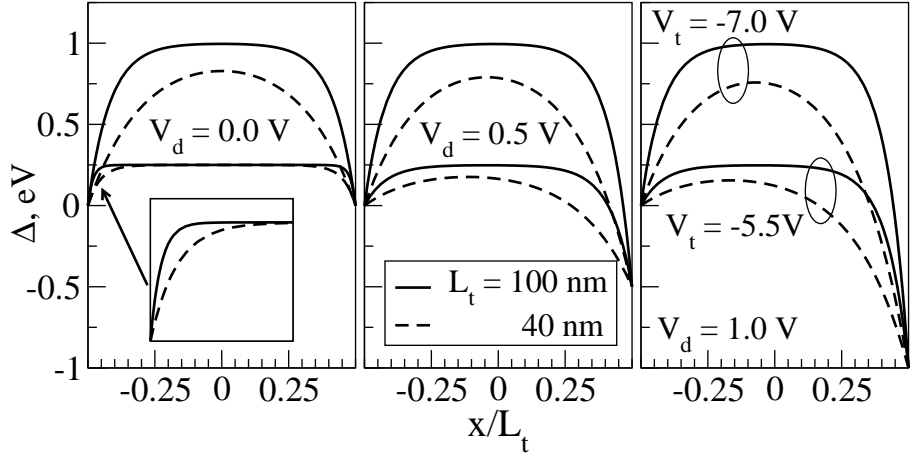}
\end{center}
\label{fig3}
\caption{Barrier profile $\Delta = -e\varphi$
at  different top-gate voltage $V_t$ and drain voltage $V_d$ for GBL-FETs with different
top-gate length $L_t$. Upper and lower pairs of curves correspond
to $V_t - V_{th} = - 1.5$~V and  $V_t - V_{th} \simeq 0 $, respectively; $W = 10$~nm$, b = 0.05$,
and $V_b = 5.0$~V.
}
\end{figure*}

At relatively high drain voltages ($eV_d > \varepsilon_F \gg k_BT$),
the electron charge in the source portion of the gated section ($x \leq x_m$,
 where $x_m$ is the coordinate of the barrier top) is
primarily determined by the electrons injected from the source. 
The electron injection from the drain at high drain voltages is insignificant.
Hence, the electron charge in the drain portion of the gated section can be disregarded.
In this case, Eq.~(2) can be presented as

\begin{equation}\label{eq40}
\frac{d^2 \varphi}{d x^2} -
\frac{\varphi}{\Lambda^2}
=  \frac{F}{\Lambda^2}
\end{equation}
at $-L_t/2 \leq x  \leq x_0$, and
\begin{equation}\label{eq41}
\frac{d^2 \varphi}{d x^2} -
\frac{\varphi}{\Lambda_0^2}
=  \frac{F_0}{\Lambda_0^2}
\end{equation}
at $x_0 \leq x \leq   L_t/2$.

At the point $x = x_0$ corresponding to the condition $e\varphi|_{x = x_0} + \varepsilon_F = 0$,
 the solutions of Eqs.~(40) and (41) should be matched: 
\begin{equation}\label{eq42}
\varphi\biggr|_{x = x_0 -0} = \varphi\biggr|_{x = x_0 +0} = - \frac{\varepsilon_F}{e}, \qquad
\frac{d\varphi}{dx}\biggr|_{x = x_0 -0} = \frac{d\varphi}{dx}\biggr|_{x = x_0 +0}.
\end{equation}

Solving Eqs.~(40) and (41) with  conditions (4) and (42),
we obtain the following formulas for the potential $\varphi$ at $-L_t/2 \leq x \leq x_0$
and $x_0 \leq x  \leq L_t/2$ as well as an equation for $x_0$:

\begin{widetext}
\begin{equation}\label{eq43}
\varphi =  F\biggl[\frac{\cosh(x/\Lambda)}{\cosh(L_t/2\Lambda)} -1\biggr] 
 -\biggl[\frac{\varepsilon_F}{e} +
 F\frac{\cosh(x_0/\Lambda)}{\cosh(L_t/2\Lambda)} - F \biggr]
\frac{\sinh[(x + L_t/2)/\Lambda)]}{\sinh[(x_0 + L_t/2)/\Lambda)]},
\end{equation}

$$
\varphi =  F_0
\biggl[\frac{\cosh(x/\Lambda_0)}
{\cosh(L_t/2\Lambda_0)} - 1 \biggr] + 
V_d\,\frac{\sinh[(x + L_t/2)/\Lambda_0]}
{\sinh(L_t/\Lambda_0)} 
$$
\begin{equation}\label{eq44}
- \biggl[\frac{\varepsilon_F}{e} +F_0\frac{\cosh(x_0/\Lambda_0)}
{\cosh(L_t/2\Lambda_0)} - F_0  + 
V_d\,\frac{\sinh[(x_0 + L_t/2)/\Lambda_0]}
{\sinh(L_t/\Lambda_0)} \biggr]\frac{\sinh[(x - L_t/2)/\Lambda_0]}{\sinh[(x_0 - L_t/2)/\Lambda_0]},
\end{equation}

$$
\frac{1}{\Lambda}\biggl\{F\frac{\sinh(x_0/\Lambda)}{\cosh(L_t/2\Lambda)} 
 - \biggl[\frac{\varepsilon_F}{e} +
F\frac{\cosh(x_0/\Lambda)}{\cosh(L_t/2\Lambda)} - F\biggr]
\frac{\cosh[(x_0 + L_t/2)/\Lambda)]}{\sinh[(x_0 + L_t/2)/\Lambda)]}\biggr\}
$$
$$
=  \frac{1}{\Lambda_0}
\biggl\{F_0\frac{\sinh(x_0/\Lambda_0)}
{\cosh(L_t/2\Lambda_0)}  + 
V_d\,\frac{\cosh[(x_0 + L_t/2)/\Lambda_0]}
{\sinh(L_t/\Lambda_0)} 
$$
\begin{equation}\label{eq45}
- \biggl[\frac{\varepsilon_F}{e} 
+ 
F_0\frac{\cosh(x_0/\Lambda_0)}{\cosh(L_t/2\Lambda_0)} - F_0  
+ 
V_d\,\frac{\sinh[(x_0 + L_t/2)/\Lambda_0]}
{\sinh(L_t/\Lambda_0)} \biggr]
\frac{\cosh[(x_0 - L_t/2)/\Lambda_0]}{\sinh[(x_0 - L_t/2)/\Lambda_0]}\biggr\}.
\end{equation}
\end{widetext}

In the cases $V_b + V_t \simeq 0$ and  
$-(V_b + V_t) \gtrsim V_b \gg \varepsilon_F/e$, Eq.~(45) yields
$x_0 = - L_t/2 + \Lambda\ln[4b V_t/(\sqrt{b}+ 2b)(V_b + V_t)]$ and 
$
x_0 \simeq - L_t/2 + 2\Lambda_0\varepsilon_F/[-e(V_b + V_t)]
\simeq - L_t/2 + 2b\Lambda_0V_b/[-(V_b + V_t)]$,
respectively.
When $-(V_b + V_t)
\rightarrow +0$, the matching point shifts toward the channel center.
If  $-(V_b + V_t)$ increases, the matching point tends to
the the source edge of the channel. In this case, the role of the electron charge
in the vicinity of the source edge diminishes, and the potential distribution 
tends to that given by Eq.~(27).

At the threshold, the matching point $x_0$ coincides with
the position of the barrier maximum $x_m$.
Considering that at $x_0 = x_m$, both the left-hand and right-hand sides of Eq.~(45)
are equal to zero, for the barrier top height near the threshold at relatively high drain voltages  we obtain the following:

\begin{equation}\label{eq46}
\Delta_m \simeq eF_0
\biggl(1 - \frac{1}{\eta_0}\biggr)
-  \frac{eV_d}{\eta_0} =  - \frac{e(V_b + V_t)}{2}
\biggl(1 - \frac{1}{\eta_0}\biggr)
-  \frac{eV_d}{\eta_0},
\end{equation}
\begin{equation}\label{eq47}
x_m \simeq - \frac{L_t}{2} + 
\Lambda\ln\biggl(\frac{2F}{F - F_0}\biggr) \simeq - \frac{L_t}{2} +
\Lambda\ln\frac{1}{b}.
\end{equation}
Both Eqs.~(29) and (46) correspond to the situations when the electron density
in a significant portion of the channel is fairly low. However there is a distinction
in the dependence of $\Delta_m$ on $V_d$ (the pertinent coefficients differ by factor of two).
This is because in the first case  the barrier top  is located near the channel center, whereas
in the second case it is shifted to the vicinity of the source edge
[compare Eqs.~(30) and (47)].

Using Eqs.~(19) and (50), we obtain  $(\partial \Delta_m/\partial V_t) = - 
(e/2)(1- \eta_0^{-1})$ and arrive at the following formula for the transconductance
near the threshold, i.e., when   $V_t \simeq V_{th}$

\begin{equation}\label{eq48}
J \simeq J_0\,\biggl[\zeta_0  + \zeta_1\frac{e(V_t - V_{th})}{2k_BT}\biggl(1 - \frac{1}{\eta_0}\biggr)
\biggr],
\end{equation}
\begin{equation}\label{eq49}
g \simeq \frac{J_0e}{k_BT} \frac{\zeta_1}{2}\biggl(1 - \frac{1}{\eta_0}\biggr) = g_{th}.
\end{equation}
Here, as above,
$V_{th} \simeq - (1 + 2b)V_b$. In particular, Eqs.~(48) and (49) at $V_t = V_{th}$,
yield $J_{th} \simeq J_0\zeta_0$.
For a GBL-FET with $L_t = 40$~nm, $W = 10$~nm, at  $T = 300$~K, one obtains 
$J_{th} \simeq1.656$~A/cm and  $g_{th} \simeq 2167$~mS/mm.

\subsection{Low top-gate  voltages ($\Delta_m < \varepsilon_F$).}

At relatively low top-gate voltages when $\Delta_m < \varepsilon_F$,
the electron system is degenerate not only in the source and drain sections but
in the gate section as well. In this top-gate voltage range,
the spatial variation of the potential is characterized by
$\Lambda \simeq  W\sqrt{2b/3}$.
As a result, for $\Delta_m$ one obtains an equation similar to Eq.~(37).
Since $\Lambda < \Lambda_0 \ll L_t$, the parameter 
determining the effect  of the top-gate geometry and 
the effect of  drain-induced barrier lowering is
$\eta = \exp (L_t/2\Lambda)/2 \gg \eta_0$. As a consequence, one can neglect
the effects in question in the  top-gate voltage  range under consideration.
In this case, $\varepsilon_F - \Delta_m \simeq bV_t$ is
As a result, one can arrive at 
\begin{equation}\label{eq50}
J \simeq \frac{2}{3}I_0e^{3/2}[b(V_t - V_{th})]^{3/2}
- 
[b(V_t - V_{th} - V_d]^{3/2}
\end{equation}
when $V_d \leq b(V_t - V_{th})$, 
\begin{equation}\label{eq51}
J \simeq \frac{2}{3}I_0e^{3/2}[b(V_t - V_{th})]^{3/2},
\end{equation}
when $V_d > b(V_t - V_{th})$, 
\begin{equation}\label{eq52}
\frac{\partial \Delta_m}{\partial V_t} = - be. 
\end{equation}
Considering Eq.~(51),
at low top-gate voltages we obtain 

\begin{equation}\label{eq53}
g \simeq \frac{beI_0}{2}
\frac{eV_d}{\sqrt{\varepsilon_F}} \simeq \frac{\sqrt{b}e^{3/2}I_0}{2}
\frac{V_d}{\sqrt{V_b}} 
\end{equation}
when  $eV_d \ll  \varepsilon_F - \Delta_m \simeq be(V_b + V_t)$, 
and
\begin{equation}\label{eq54}
g \lesssim  beI_0\,
\sqrt{\varepsilon_F} \simeq  b^{3/2}e^{3/2}I_0\,
\sqrt{V_b} = g_{on}
\end{equation}
when $eV_d \gg  \varepsilon_F - \Delta_m \simeq be(V_b + V_t)$.
As follows from Eqs.~(33) and (34),
the transconductance is proportional to a small parameter $b^{3/2}$. This is because
the effect of the top-gate potential is weakened due  to
a strong screening by the degenerate electron system in the gated section.
As a result, the transconductance at low top-gate voltages is smaller than that
at the top-gate voltage corresponding to the threshold.
Assuming that $b = 0.05$ and $V_b = 5$ ($\varepsilon_F \simeq 0.25$~eV), 
 we obtain $g_{on} \simeq 1467$~mS/mm. Comparing Eqs.~(49) and (54), we find
$g_{on}/g_{th} \propto b^{3/2}\sqrt{eV_b/k_BT} \simeq 0.158$ and 
 $g/g_{th}\lesssim  g_{on}/g_{th}\simeq 0.68$.

Since the source-drain current at high top-gate voltages decreases exponentially
when $-V_t$ increases, the transconductance decreases as well.
 
Figure~3 shows the barrier (conduction band)  profile $\Delta = -e\varphi$ 
in the GBL-FETs calculated
using Eqs.~(27), (36), and (44) for different applied voltages and top-gate lengths.
As demonstrated, the barrier heigh naturally decreases with increasing
$V_t - V_{th}$. At the threshold ($V_t = V_{th}$),
the barrier heigh is equal to the Fermi energy (at $b = 0.05$ and
$V_b = 5$~V, $\varepsilon_F \simeq 0.25$~eV).
One can see that shortening of the top-gate leads to
a marked decrease in the barrier height (the short-gate effect). 
The source-drain current as a function of the drain voltage 
for different top-gate voltage swings is demonstrated in Fig.~4.
The dependences corresponding to $V_t - V_{th} < 0$,  $V_t - V_{th} = 0$,
and  $V_t - V_{th} > 0$ were calculated using formulas from subsections A, B, and C, respectively.
Figure~5 shows that the transconductance as a function of the top-gate voltage swing
exhibits a pronounced maximum at $V_t \simeq V_{th}$.
This maximum is attributed to the following.
At high top-gate voltages, the effect of screening  is insignificant due to low electron density in the channel. As a result, the height of the barrier top is rather sensitive to the top-gate voltage variations. The source-drain current in this case is exponentially small,
so that the transconductance is small.
In contrast, at low top-gate voltages, the screening by the electrons in the channel is 
effective, leading to a much weaker control of the barrier height by the top voltage
[pay attention to parameter $b \ll 1$ in Eqs.~(50) - (54)]. Despite, a strong source-drain current provides a moderate values of the transconductance.
However, in the near threshold voltage range, both the sensitivity of the barrier height to the top-gate volatage and the source-drain current are  fairly large.

\begin{figure}[t]
\begin{center}
\includegraphics[width=7.0cm]{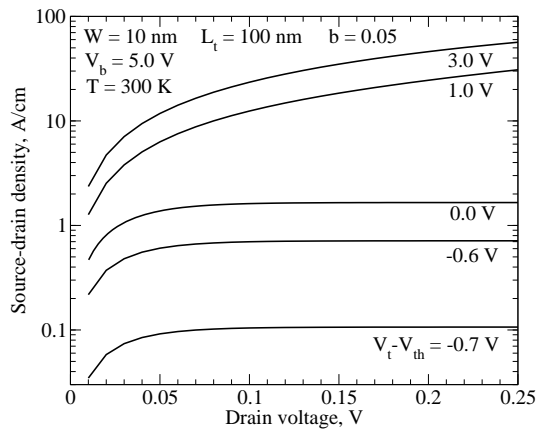}
\end{center}
\label{fig4}
\caption{Source-drain current $J$ versus drain voltage $V_d$
 at different values of the
top-gate voltage swing $V_t - V_{th}$.
}
\end{figure}

\begin{figure}[t]
\begin{center}
\includegraphics[width=7.0cm]{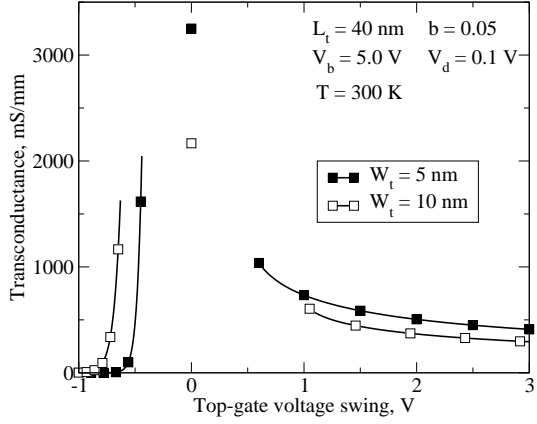}
\end{center}
\label{fig5}
\caption{Transconductance $g$ versus top-gate voltage swing $V_t - V_{th}$.
}
\end{figure}


\section{Discussion}

\subsection{Role of geometrical parameters}

In the main part of the paper, we assumed that the thicknesses of the gate layers are equal
to each other: $W_b = W_t = W$.
If  $W_b \neq W_t$, the formulas obtained above should be  slightly modified.
In particular, near the threshold $(\partial \Delta_m/\partial V_t) = - 
(e/(1 + W_t/W_b))(1- \eta_0^{-1})$, so that the transconductance instead of Eq.~(49)
is given by 
\begin{figure}[t]
\begin{center}
\includegraphics[width=7.0cm]{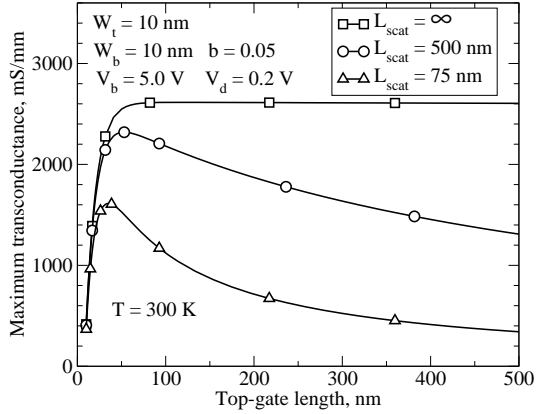}
\end{center}
\label{fig6}
\caption{Maximum transconductance $g$ versus top-gate length $L_t $ for $W_t = 10$~nm.
}
\end{figure}
\begin{figure}[t]
\begin{center}
\includegraphics[width=7.0cm]{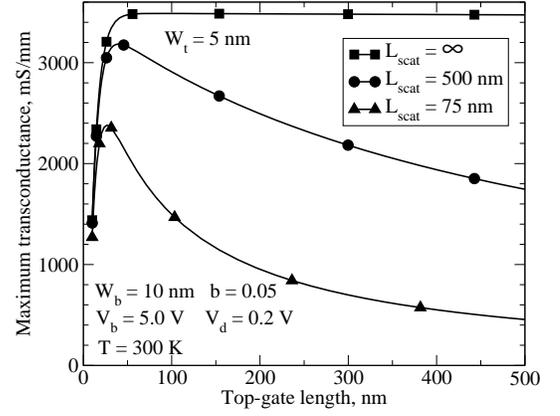}
\end{center}
\label{fig7}
\caption{The same as in Fig.~6 but for $W_t = 5$~nm.
}
\end{figure}
%
$$
g \simeq \frac{J_0e}{k_BT} \frac{\zeta_1}{(1 + W_t/W_b)\zeta_0}
\biggl(1 - \frac{1}{\eta_0}\biggr)
$$
\begin{equation}\label{eq55}
\simeq \frac{J_0e}{k_BT} \frac{\zeta_1}{(1 + W_t/W_b)\zeta_0}.
\end{equation}
As follows from the comparison of Eqs.~(49) and (55), changing $W_t$, in particular,
 from
$W_t = W_b$ to $W_t = W_b/2$, leads to an increase in the transconductance at the top-gate voltages near the threshold of 50~percent (compare the $g$ versus $V_t - V_{th}$
dependences in Fig.~5 for $W_b = W_t = 10$~nm and $W_b = 10$~nm and $W_t = 5$~nm. 

As demonstrated above,  shortening of the  top gate can result in deterioration of the
GBL-FET characteristics. This effect 
is characterized by parameters $\eta$ and $\eta_0$, which strongly depends 
on the top-gate length $L_t$. This effect gives rise to a decrease in the transconductance
if $L_t$ becomes smaller.
However, in the range of large $L_t$ the electron collisions
can play a substantial role. This leads to the transconductance  roll-off with increasing $L_t$. The effect of  drain-induced barrier lowering
is also characterized by  parameters $\eta$ and $\eta_0$.
As follows from Sec.~IV, at sub-threshold, 
 the effect of drain-induced barrier lowering results in the appearance of the factor
$\exp(eV_d/2\eta_0k_BT)$ [see Eqs.~(31) and (32)]. This factor can provide a marked
increase in $J$ and $g$  with increasing drain voltage in GBL-FETs with relatively short
top gates. For example, at  $W = 40$~nm, $L_t = 40$~nm,  $T = 300$~K, and $V_d = 0.25$~V,
this factor is about 2.37.

\subsection{Effect of electron scattering}

As shown above, the potential distributions in the main part of the gated section are
fairly flat. This implies that to determine the effect of electron scattering associated with  disorder
and acoustic phonons one can use Eq.~(2) with the collisional term following the approach
applied in Ref.~\cite{17}.
Considering here the case when the elastic scattering mechanisms under consideration are
strong, so that they lead to an effective isotropization of the electron distribution,
one can find that the values of the source-drain current and the transconductance obtained
in the previous section for the ballistic transport should be multiplied
by a collision factor $C$. This factor is equals to $C_{\infty} = \sqrt{2\pi\,k_BT/m}/L_t\nu$,
where $\nu = mw/2$ is the collision frequency (we put $w(q) = w = const$.
It characterizes the fraction of the electrons injected into the gated section
and those reflected back due to the collisions.
To obtain the GBL-FET characteristics with
 the  top-gate lengths in a wide range (to follow the transition
from  the ballistic electron transport  to the collision-dominated transport), we use 
for the collision factor the following interpolation formula: 

\begin{equation}\label{eq56}
C = \frac{1}{1 + L_t/L_{scat}},
\end{equation}
where $L_{scat} = \sqrt{2\pi\,k_BT/m}/\nu$ is the characteristic scattering length.
Figures~6 and 7 show the dependences of the transconductance maximum 
(approximately at $V_t = V_{th}$) on the top-gate length calculated for GBL-FETs
with different $W_t$. The scattering length is assumed to be $\infty$ (ballistic transport),
 $500$~nm, and $75$~nm.
At $T = 300$~K,
this corresponds to the collision frequencies $\nu = 0$,
$\nu \simeq 1.14\times 10^{12}$~s$^{-1}$ (the electron mobilities $\mu \simeq    1.75\times 10^{5}$~cm$^{2}$/V s) 
and  $\nu \simeq 7.6\times 10^{12}$~s$^{-1}$ ($\mu \simeq    2.63\times 10^{4}$~cm$^{2}$/V s)
, respectively).
One can see that in the case of essential electron collisions,
$g$ versus $L_t$ dependences exhibit pronounced
maxima. This is attributed to an interplay of two effects: the short-gate effect (weakening
of the barrier controllability by the top-gate voltage when $L_t$ decreases) and the effect of collisions, which reinforces when $L_t$ becomes larger.
(which decreases the current).   
As follows from Figs.~6 and 7, the electron collisions can lead to a dramatic decrease in
the transconductance.

\subsection{Charge inversion in the gated section.}

At sufficiently high top-gate voltages $V_t < V_{in}$
when $\Delta_m >  \varepsilon_F + E_g$,
the top of the valence band in the gated section of the channel can be markedly populated
by holes (inversion of the gated section charge), so that the term in the right-hand side of Eq.~(2) becomes negative.
The latter inequality corresponds to the following value of the inversion voltage
[see Eqs.(A4) and (A5)]:
\begin{equation}\label{eq57}
V_{in} = -V_b\biggl(1 + 2b + \frac{d_0}{W}\biggr) \simeq V_{th} 
\biggl(1  + \frac{d_0}{W}\biggr).
\end{equation}

The hole charge provides an effective screening of the transverse electric field
in the gated section. This leads to weakening of the sensitivity of the barrier height
and  the source-drain current on the top gate voltage $V_t$~\cite{15} and, hence, to
 a decrease in the transconductance. 
This pattern is valid at the dc voltages or when 
the characteristic time of their variation is long in comparison with
the characteristic times of the thermogeneration of holes
and the tunneling between the channel side regions
and the gated section~\cite{15,17}. In the situation when the hole recharging of the gated section of the channel is a relatively slow process, the ac transconductance at the frequencies 
higher than  the characteristic recharging  frequency can substantially exceed the dc transconductance.

\subsection{Interband tunneling}

At elevated top-gate voltages ($V_t < V_{in}$), the interband tunneling of electrons from
the conduction band in
the source to the valence band in the gated section as well as from the valence band
in the gated section to the conduction band in the drain can be essential. 
At high drain voltages the latter tunneling processes can be particularly pronounced.
This  can result in elevated source-drain current despite rather high barrier.
To limit the tunneling, the back-gate voltage, which mainly determines the energy gap,
 should be high enough. A decrease in the gate layer thicknesses  also promotes 
 the tunneling suppression.
For example, if $W = (5 - 10)$~nm,
$V_b = - V_t = 5$~V, the energy gap in the gated section of the channel
 $E_g \simeq 0.17 - 0.34$~eV [see Eq.~(A4)]. 
Despite some attempts to calculate the tunneling currents
in G-FETs  and GBL-FETs (see, for instance, Refs.~\cite{11,19}),
the problem for GBL-FETs remains open. 
This is because the spatial nonuniformity of the energy
gap in the channel and its nonlinear spatial dependence (particularly near the drain edge)
associated with the features of 
 the potential distribution
  under the applied voltages. Generalizing Eq.~(A4),
we obtain $E_g = ed[V_b - \varphi(x)]/2W$, hence the energy gap
varies from  $E_g = E_{g,s}\simeq edV_b/W $ at $x = -L_t/2$ to  $E_g = E_{g,s} \simeq  ed(V_b - V_d/2W$ at $x = L_t/2$. It reaches a maximum $E_g = d(eV_b + \Delta_m)/2W$
at $x = x_m$.
 One can see that  $E_{g,d}$ can be markedly
smaller than $ E_g$, especially at not too small drain voltages.
Due to this, the deliberation of the tunneling in GBL-FETs  requires sufficiently rigorous
device model (which could include the formulas for the potential distribution 
obtained above) 
and numerical approach.

\section{Conclusions}
We demonstrated that the developed device model of  allows 
to derive the GBL-FET characteristics:
the potential distributions along the channel and
the dependences of the source-drain current and the transconductance 
on the applied voltages and the geometrical parameters
as closed-form analytical expressions. 
The key element of the model, which provide an opportunity to solve the problem analytically,
 is the use of the Poisson equation in the
weak nonlocality approximation. In particular, the model accounts for
the effect of screening of the transverse electric field
by the electron charge in the channel, the short-gate effect, and
the effect of  drain-induced barrier lowering.
The parameters $\eta_0$ and $\eta$ characterizing the strength of these effects
in the cases of essentially depleted channel and strong screening were expressed
via the geometrical parameters and the Bohr radius.
As shown,  the GBL-FET transconductance exhibits a pronounced
maximum as a function of the top-gate voltage swing. The interplay of
 the short-gate effect
and the electron collisions results in a nonmonotonic dependence of the transconductance
on the top-gate length. 
The obtained analytical
formulas for the potential barrier height, the source-drain current, and the transconductance
can be used for GBL-FET optimization by proper choice of the thicknesses of gate layers,
the top-gate length, and the bias voltages.

\section*{Acknowledgments}
The authors are grateful to H.~Watanabe for stimulating comments
and to V.~V.~V'yurkov for providing Ref.~\cite{31}.
The work was supported by the Japan Science and Technology Agency, CREST,
Japan.

 \section*{Appendix}
\setcounter{equation}{0}
\renewcommand{\theequation} {A\arabic{equation}}

Disregarding the effect of ``Mexican-hat'' and the nonparabolicity of
the electron energy spectrum,  the density of states can be considered independent of the energy (in the energy range under consideration). Taking this into account,
the electron Fermi energies  and the energy gaps
in the source and drain  sections of
the GBL-FET channel are, respectively, given by~\cite{17,18} ($W_b = W_t = W$)
\begin{equation}\label{eqA1}
\varepsilon_{F,s} = \frac{k_BT}{(1 + b)}
\ln\biggl[\exp\biggl(\frac{beV_b}{k_BT}\biggr) - 1\biggr],
\end{equation}
\begin{equation}\label{eqA2}
\varepsilon_{F,d} = \frac{k_BT}{(1 + b)}
\ln\biggl[\exp\bigg(\frac{be(V_b - V_d)}{k_BT}
\biggr) - 1\bigg],
\end{equation}
\begin{equation}\label{eqA3}
E_{g,s} = \frac{edV_b}{2W}, \qquad E_{g,s} = \frac{ed(V_b -V_d)}{2W}.
\end{equation}
Here $a_B = k\hbar^2/me^2$, $b = a_B/8W$, and $d \lesssim d_0$, 
where $d_0 \simeq 0.34$~nm~\cite{30} is the spacing between 
the graphene layers in the GBL, while $d$ stands for the effective spacing accounting for the screening of the transverse electric field by GBL (polarization effect).
In portion of the gated section essentially occupied by electrons and 
its depleted portion, one obtains
\begin{equation}\label{eqA4}
E_{g} = \frac{ed(V_b - V_t)}{2W}, \qquad E_{g} = \frac{ed_0(V_b - V_t)}{2W},
\end{equation}
respectively. 
Due to $a_B \gg d$, from Eqs.~(A1) - (A4) one obtains 
$\varepsilon_{F,s} \geq
\varepsilon_{F,d} > E_{g,s} \geq  E_{g,d}$.
At $V_t < 0$, $E_g$ can significantly exceed $ E_{g,s}$ and $E_{g,d}$.
In the case of strong degeneracy of the electron system,
Eqs.~(A1) and (A2) yield
\begin{equation}\label{eqA5}
\varepsilon_{F,s} \simeq eV_b\frac{a_B/8W}{(1 + b)},
\qquad
\varepsilon_{F,d} \simeq e(V_b - V_d)\frac{b}{(1 + b)}.
\end{equation}
The quantity $\varepsilon_{F,d}$ is given by the same equations in which,
however, $V_b$ is substituted by $V_b - V_d$.
As a result, 
\begin{equation}\label{eqA6}
\varepsilon_{F,s} - \varepsilon_{F,d} \simeq \frac{beV_d}
{1 + b}
\simeq beV_d,
\end{equation}
\begin{equation}\label{eqA7}
eV_d^* = eV_d + \varepsilon_{F,d} - \varepsilon_{F,s} 
\simeq \frac{eV_d}{1 + b}.
\end{equation}
Since parameter $b$ in reality is small, so that 
$\varepsilon_{F,s} - \varepsilon_{F,d} \simeq b eV_d \ll eV_d$ and 
$V_d^* \simeq V_d$
we put $\varepsilon_{F,s} = \varepsilon_{F,d} = \varepsilon_F$ 
and substitute $V_d^*$ by $V_d$.

%



\begin{thebibliography}{30}
 
\bibitem{1}
C. Berger, Z. Song, T. Li, X. Li,  A.~Y.~Ogbazhi,R.~ Feng,
Z.~Dai,  A.~ N.~Marchenkov,  E.~ H.~Conrad,  P.~N.~First, and  W.~A.~de~Heer,
J.~Phys. Chem. {\bf 108}, 19912 (2004). 
%

\bibitem{2}
 K. S.~Novoselov,  A. K.~Geim,  S. V.~Morozov, D.~Jiang,
 M.~I.~Katsnelson,  I.~V.~Grigorieva,  S.~V.~Dubonos, and 
 A. A.~Firsov,
Nature {\bf 438}, 197 (2005).

\bibitem{3}
 A.~H.~Castro Neto, F. Guinea,    N.~ M.~ R.~Peres, K.~S.~Novoselov, and 
 A.~K.~Geim,
Rev. Mod. Phys. {\bf 81}, (2009). 
\bibitem{4}
J.~Bai, X.~Zhong, S.~Jiang, Y.~Huang, and X.~Duan,
Nature Nanotechnology {\bf 5}, 190 (2010).

\bibitem{5}
V.~Ryzhii, M.~Ryzhii, and T.~Otsuji,
J. Appl. Phys. {\bf 101}, 083114 (2007). 
\bibitem{6}
F.~Rana,  IEEE Tran. Nanotechnol. {\bf 7}, 91 (2008).
\bibitem{7}
F.~Xia, T.~Murleer, Y.-M.~Lin, A.~Valdes-Garsia,
and P.~Avouris, Nat. Nanotechnol. {\bf 4}, 839 (2009).
\bibitem{8}
V.~Ryzhii, M.~Ryzhii, V.~Mitin, and T.~Otsuji,
J. Appl. Phys. {\bf 107}, 054512 (2010). 

\bibitem{9}
V.~Ryzhii, A.~A.~Dubinov, T.~Otsuji, V.~Mitin, and M.~S.~Shur, 
J. Appl. Phys. {\bf 107}, 054505 (2010).

\bibitem{10}
V.~Ryzhii, M.~Ryzhii, and T.~Otsuji, 
Appl. Phys. Express {\bf 1}, 013001 (2008).

\bibitem{11}
V.~Ryzhii, M.~Ryzhii, and T.~Otsuji,
Phys. Status Solidi (a) {\bf 205}, 1527 (2008).

\bibitem{12}
Y.~Quyang, Y.~Yoon, J.~K.~Fodor, and  J.~Guo,
Appl. Phys. Lett. {\bf 89} 203107 (2006). 

\bibitem{13}
G.~Fiore  and G.~Iannaccone, IEEE Electron Device Lett. {\bf 28},
 760 (2007).

\bibitem{14}
G.~Liang, N.~Neophytou, D.~E.~Nikonov, and M.~S.~Lundstrom,
 IEEE Trans. Electron Devices {\bf 54}, 677 (2007).

\bibitem{15}
V.~Ryzhii,  M.~Ryzhii, A.~Satou, and T.~Otsuji,
J. Appl. Phys.  {\bf 103}, 094510 (2008).
\bibitem{16}
M.~Ryzhii, A.~Satou,  V.~Ryzhii, and T.~Otsuji,
J. Appl. Phys.  {\bf 104}, 114505 (2008).

\bibitem{17}
V.~Ryzhii,   M.~Ryzhii,  A.~Satou, T.~Otsuji, and N.~Kirova, 
J. Appl. Phys.  {\bf 105}, 104510 (2009).

\bibitem{18}
M.~Ryzhii and V.~Ryzhii,  Phys. Rev. B {\bf 79}, 245311 (2009). 
\bibitem{19}
M.~Cheli, G.~Fiori, and G.~Iannaccone,
IEEE Trans. Electron Devices {\bf 56}, 2979 (2009).

\bibitem{20}
S.~O.~Koswatta, S.~Hasan, M.~S.~Lundstrom, M.~P.~Anantram, and D.~E.~Nikonov, 
Appl. Phys. Lett. {\bf 89}, 023125 (2006).
\bibitem{21}
M.~Lenzi, P.~Palestri, E.~Gnani, S.~Reggiani, A.~Gnudi, D.~Esseni, L.~Selmi, and 
G.~Baccarani,
 Trans. Electron Devices {\bf 55}, 2087 (2008).

\bibitem{22}
S. Fregonese, J.~Gouet, C.~Manex, and T.~Zimmer,   
IEEE Trans. Electron Devices {\bf 56}, 1184 (2009).

\bibitem{23}
K. Natori, J.Appl. Phys. {\bf 76}, 4881 (1994).

\bibitem{24}
A.~Rahman, J.~Guo, S.~Datta, and M.~S.~Lundstrom,  Trans. Electron Devices {\bf 50}, 1853 (2003).
\bibitem{25}
F.~G.~Pikus and K.~K.~Likharev,
Appl. Phys. Lett. {\bf 71}, 3661 (1997).
\bibitem{27}
V.~A.~Sverdlov,T.~J~Walls, and K.~K.~Likharev, 
Trans. Electron Devices {\bf 50}, 1926 (2003).%
\bibitem{27}
N.~Sano,  Phys. Rev. Lett. {\bf 93}, 246803 (2004).
\bibitem{28}
G.~Mugnaini  and G.~Iannaccone, Trans. Electron Devices {\bf 52},
 1802 (2005).

\bibitem{29} 
R.~Kim,  P.~A.~Neophytou, G.~ Klimeck, and M.~S.~Lundstrom, 
J. Vac. Sci. Technol. {\bf B26}, 1628 (2008). 

\bibitem{30}
R.~Akis, N.~Faralli, D.~K.~ Ferry, S.~M.~Goodnick, K.~A.~Phatak, 
and M.~Saratini,  
IEEE Trans.  Electron Devices {\bf 56}, 2935 (2009).
\bibitem{31}A.~N.~Khomyakov and V.~V.~V'yurkov,
Russian Microelectronics, {\bf 38}, 393 (2009).


\bibitem{32}
T.~Ohta, A.~Q.~Bostwick, T.~Seyller, K.~Horn, and E.~Rotenberg,  
Science {\bf 333}, 951 (2006).

\bibitem{33}
E.~McCann, Phys. Rev. B {\bf 74}, 161403 (2006).

\bibitem{34} 
E.~V.~Castro, K.~S.~Novoselov, S.~V.~Morozov, N.~M.~R.~Peres, 
 J.~M.~ B.~Lopes dos Santos,
L.~Nilsson, F.~Guinea, A.~K.~Geim, and A.~H.~Castro Neto, J. Phys.: Condens. Matter
{\bf 22}, 175503 (2010).

\bibitem{35}
A.~A.~Sukhanov  and Y.~Y.~Tkach, Sov. Phys. Semicond {\bf 18} 797 (1984).
\bibitem{36}
A.~O.~Govorov, V.~M.~Kowalev, and A.~V.~Chaplik, 
JETP Lett. {\bf 70}, 488 (1999).

\bibitem{37}
S. ~M.~Sze, {\em Physics of Semiconductor Devices}
(Wiley, New York, 1981).
\bibitem{38}
M.~Shur, {\em Physics of Semiconductor Devices} (Prentice-Hall, New Jersey, 1990). 

\end{thebibliography}
\end{document}